\newcount\eqnno\eqnno=0\newcount\secn\secn=0
\def\numeqn{\global\advance\eqnno by 1 \eqno(\the\eqnno)}
\font\bigbf=cmbx10 scaled\magstep2
\font\bigbish=cmbx10 scaled\magstep1
\def\Stephen{9}
\def\hillwidrow{10}
\def\us2{11}

\magnification=\magstep1
\pageno=0\nopagenumbers
{\rightline{DAMTP/96-73}}
\rightline{SWAT/132}
\bigskip\bigskip
\centerline{\bigbf  DISSIPATING COSMIC VORTONS}
\vskip 9pt
\centerline{\bigbf  AND BARYOGENESIS}
\vskip 1 true in
\centerline{\bigbish Anne-Christine Davis$^1$}
\centerline{\bigbish Warren B. Perkins$^2$}
\bigskip
\centerline{1.\ \it Department of Applied Mathematics and Theoretical Physics,}
\centerline{\it Silver Street, Cambridge, CB3 9EW, Great Britain}
\centerline{2.\ \it Department of Physics, University of Wales Swansea,}
\centerline{\it Singleton Park, Swansea, SA2 8PP}
\vskip 1 true in
\centerline{\hfill{\bf ABSTRACT}\hfill}\nobreak\smallskip\noindent

Grand unified theories can admit cosmic strings with fermion zero modes
which result in the string carrying a current and the formation
of stable remnants, vortons.
We consider  theories in which the zero modes do not survive
a subsequent phase transition, for example the electroweak transition,
 resulting in vorton dissipation. The dissipating vortons can create
a baryon asymmetry. We calculate the asymmetry produced, and show that
it is maximised if the vortons decay just before they dominate the energy
density of the Universe. We further bound the asymmetry produced by 
the late decay of any relic particle.

\def\startnumberingat#1{\pageno=#1\footline={\hss\tenrm\folio\hss}}
\vfill\eject
\startnumberingat1

\leftline {\bf 1) Introduction}
\overfullrule=0pt
\smallskip
The rich microstructure of cosmic strings is starting to receive considerable
attention.  In
particular, the additional features acquired by the defect core 
 at each
subsequent symmetry breaking [1]. The microstruture of the strings has
been used to constrain general particle physics theories to ensure that
they are consistent with standard cosmology [2]. In particular, if a
theory admits cosmic strings which subsequently become superconducting,
then an initially weak current on a closed string loop will amplify as
the loop contracts. The current may become sufficiently strong to halt
the contraction of the loop, preventing it from decaying. A stable state,
or vorton [3], is formed. The density of vortons is tightly constrained
by the requirement that they do not over close the universe. This has
been used in [2] to constrain such models.

Strings can become superconducting due to boson condensates or to
fermion zero modes. The resulting vorton is classically stable [4].
The quantum stability is an open question. It has been assumed that,
if vortons decay, they do so by quantum mechanical tunnelling. This
would result in them being very long lived. However, in the case of
fermion superconductivity, the existence of fermion zero modes at
high energy does not guarantee that such modes survive subsequent
phase transitions. The disappearance of such zero modes could give
another channel for the resulting vortons to decay.

For example, one popular particle physics theory which admits cosmic strings is
that based on the grand unified group SO(10). The resulting strings
have right handed neutrino, $\nu_R$, zero modes. The presence of zero modes means that the
string is superconducting, albeit with a neutral current [5]. The current
on the string starts as a small random current, resulting from string
inter-commuting and self-intersection. If the current is sufficiently
strong then vortons will  form. However, due to the way
the Higg's field couples, the $\nu_R$ zero mode ceases to be a zero
mode of the theory after the electroweak phase transition[9]. 
This results in the string current discharging over a finite time. Thus 
vortons formed before the electroweak phase transition will decay. 

Since the underlying theory is a grand unified SO(10) theory, and the
full SO(10) symmetry is restored in the core of the string, there
will be SO(10) gauge bosons in the string core. When the vorton decays
these are released and their out-of-equilibrium decay results in a baryon
asymmetry being produced. This is similar to that produced by collapsing
string loops [6] or by monopole annihilation [7]. 

In this letter we address this problem. First we review  neutrino currents
and the vorton density produced and
 then calculate the  baryogensis resulting 
from  vorton decay and the change in entropy density. For vortons decaying at the electroweak 
transition the baryon asymmetry
produced by this mechanism is not sufficient to account for nucleosynthesis.
However, in theories with an intermediate phase transition such a mechanism
may give the required baryon asymmetry. We compare our results with a general  
 limit on baryogenesis produced by the late decay of a relic particle.

\vskip 15pt

\leftline {\bf 2. Fermion Zero Modes and Vortons}
\smallskip
The SO(10) string has $\nu_R$ zero modes [5], the presence of which
result in the string being superconducting. 
Despite being electrically neutral, this still applies to the $\nu_R$,
the possibility of neutral current carriers in the string was first developed
in [8]. The string can build up a random current, similar to that
in the bosonic case, resulting from string self-intersections and 
intercommuting. When the string 
self-intersects or intercommutes there is a finite probability that
the fermi levels will be excited. This produces a distortion in the
fermi levels, resulting in a current flow, similar to that discussed
by Witten [5]. This results in a smaller current than with an external 
magnetic field, but the current is still sizable for a grand unified string.

For strings that are formed at a temperature $T_{\rm x}$  and become superconducting at 
formation, the vorton number density is given by[2]
$$
n_v=\nu_* f \bigl({\beta T_{\rm x} \over m_{\rm Pl}}\bigr)^{3\over 2} T^3,
$$
while the vorton mass density is
$$
\rho_v=\nu_* f \bigl({\beta T_{\rm x} \over m_{\rm Pl}}\bigr)^{5\over 4}
T_{\rm x} T^3,
$$
where $\nu_*$, $f$ and $\beta$ are factors of order unity.

The neutrino zero modes in SO(10) GUT do not remain zero modes after the
electroweak phase transition [\Stephen], becoming instead  low lying
bound states. While bound states can carry a current, this current is
transient[\hillwidrow]. As the current decays, angular momentum is lost,
the vortons shrink and eventually decay. The details of the decay process 
are discussed in ref.\us2. As the vortons decay the GUT particles they hold 
are released and then themselves decay. The baryon asymmetry produced by
these decaying particles is discussed in the next section.

\vskip 15pt
\leftline {\bf 3. Baryogenesis from Vortons}
\smallskip
Given the number density of vortons at the electroweak phase transition we can 
estimate the baryon asymmetry produced by vorton decay using,
$$
{n_b \over s}={n_v \over s}\epsilon N,
$$
where $s$ is the entropy density, $\epsilon$ is the baryon asymmetry produced by a 
single SO(10) GUT particle and N is the number of
GUT particles per vorton. We need to consider two cases: firstly the vortons may decay before they dominate the
energy density of the universe and we do not need to know the time scale for vorton decay since $n_v/s$ is
 an invariant quantity. Alternatively, if the vorton energy density does dominate the energy density of the 
Universe we must modify the temperature evolution of the Universe to allow for entropy generation.

Assuming that the Universe is radiation dominated until after the electroweak phase transition, the temperature of the 
Universe is simply that of the standard hot big bang. We can estimate the entropy density following vorton decay
using the standard result,
$$
s={2\pi^2 \over 45} g^* T^3,
$$
where $g^*$ is the effective number of degrees of freedom at the electroweak scale ($\simeq 100$).
The vorton to entropy ratio is then
$$
{n_v \over s} \simeq \bigl({T_{\rm x}\over m_{\rm Pl}}\bigr)^{3\over 2}{45\over 2\pi^2 g^*}
\sim 5\times 10^{-6},
$$
for $T_{\rm x}\sim 10^{16}$GeV.

The number of GUT particles per vorton is given by[2]
$$
N=\bigl({\beta T_{\rm x} \over m_{\rm Pl}}\bigr)^{-{1\over 4}}
\sim 10,
$$
and we have 
$$
{n_b \over s}\sim 10^{-5}\epsilon.
$$

Alternatively, the vorton energy density may come to dominate and we must allow for a non-standard temperature evolution.
The temperature of vorton-radiation equality, $T_{\rm veq}$, is given by
$$
T_{\rm veq}={\nu_* f\over g^*} \bigl({\beta T_{\rm x} \over m_{\rm Pl}}\bigr)^{5\over 4}T_{\rm x}.
$$
If we assume that the vortons decay at some temperature $T_d$ and reheat the Universe to a temperature $T_{\rm rh}$, we
 have
$$
\hat g^*  T_{\rm rh}^4=\rho_v(T=T_{\rm d})=
\nu_* f \bigl({\beta T_{\rm x} \over m_{\rm Pl}}\bigr)^{5\over 4}
T_{\rm x} T_{\rm d}^3,
$$
where $\hat g^*$ is the number of degrees of freedom for this lower temperature. We then have,
$$
{T_{\rm rh}\over T_{\rm eq}}
 =[{g^*\over \hat g^*} \bigl({T_{\rm d} \over T_{\rm eq}}\bigr)^3]^{1\over4 }
\qquad
{\rm or}
\qquad
{T_{\rm rh}\over T_{\rm d}}=[{g^*\over \hat g^*}{ T_{\rm eq}\over T_{\rm d}}] ^{1\over4 }.
$$
This reheating and  entropy generation leads to an extra baryon dilution factor of
$$
({T_{\rm rh}\over T_{\rm d}})^{-3}=[{g^*\over \hat g^*}{ T_{\rm eq}\over T_{\rm d}}] ^{-{3\over 4}}.
$$
In this case the baryon asymmetry produced by the decaying vortons is given by
$$
{n_b\over s}={n_v\over s} N\epsilon[{g^*\over \hat g^*}{ T_{\rm eq}\over T_{\rm d}}] ^{-{3\over 4}},
$$
where the entropy, $s$,  is that of the standard big bang model. The Universe now evolves as in the standard big bang model and 
$n_b/s$ remains invariant. Using the above results the asymmetry becomes,

$$
{n_b\over s}=\epsilon(\nu^* f {\hat g^{*3}\over g^{*'4}})^{1\over4}\beta^{5\over16}
\bigl({ T_d^{12}\over m_{\rm Pl}^5 T_{\rm x}^7}\bigr)^{1\over16}.
$$

This form is valid if the vortons dominate the energy density of the Universe before they decay, if this is not the case the dilution
factor is absent and we have
$$
{n_b\over s}\simeq{\epsilon\over g^{*'}}\bigl({ T_{\rm x}\over m_{\rm Pl}}\bigr)^{5\over4} ,
$$
as above.

For a fixed decay temperature, $T_{\rm d}$, there is a critical formation temperature above which the vortons dominate the energy density
of the Universe before they decay. For very high formation temperatures the entropy generated by the decaying vortons washes out the
baryon asymmetry, while for low formation temperatures there are few vortons and little baryon asymmetry is created. Although, in this case a sizeable 
asymmetry can be generated by loop collapse before vortons are formed [6].
The
largest baryon asymmetry is produced if the vortons decay just as they come to dominate the energy density of the Universe. This
requires a formation temperature,
$$
T_{\rm x}= [{g^*\over \nu_* f} ({m_{\rm Pl}\over\beta})^{5\over4} T_{\rm d}]^{4\over9},
$$
and produces the maximal  asymmetry given by,
$$
{n_b\over s}
\sim {\epsilon\over g^{*{4\over9}}}({T_{\rm d}\over m_{\rm Pl}})^{5\over9}.
$$
Taking the most efficient baryon asymmetry generating factor, $\epsilon\sim 0.01$, we find that in order
to generate the observed baryon asymmetry we require $T_{\rm d}>\sim 10^{-13}m_{\rm Pl}\sim
10^6GeV$. Thus vortons decaying at the electroweak scale cannot produce the observed baryon asymmetry, but
decaying vortons could generate significant baryon asymmetry if they became unstable at a sufficiently high 
energy.

We can compare these results with those 
 from a general model of baryogenesis resulting from the delayed decay of a particle. Let the
particles have mass $m$. The density of these particles once they have frozen out will be given by
$$
\rho=\alpha T^3,
$$
where $\alpha$ is a constant assuming there is no entropy generation. 
For $T\gg m$ the particles are relativistic and, assuming
that they are in equilibrium, their number density equals the number density of photons. For $T<m$ the particle number cannot
exceed the photon number as we have annihilation processes, but no production processes and we have the bound  $\alpha< m$.

If these particles never dominate the energy density, the baryon asymmetry they produce when they decay is not diluted and
is simply given by
$$
{n_b \over s}={n\over s}\epsilon \sim {\epsilon \alpha\over g^* m},
$$
where $n$ is the number density of these particles and $\epsilon$ is the baryon asymmetry produced per particle.
However, the particles will dominate the energy density of the Universe if the decay temperature, $T_{\rm d}$, is less than
the temperature of particle-radiation equality,  $T_{\rm eq} =\alpha/g^*$. As above we have a reheat temperature,
$T_{\rm rh}=(\alpha/g^{*'})^{1\over4} T_{\rm d}^{3\over4}$, which introduces a dilution factor and gives a diluted
baryon asymmetry,
$$
{n_b \over s}\sim  {\epsilon \alpha\over g^* m}\bigl({T_{\rm d}\over T_{\rm rh}}\bigr)^3
\sim \epsilon \bigl({\alpha \over g^{*'}}\bigr)^{1\over 4}{T_{\rm d}^{3\over 4}\over m}
\leq {\epsilon \alpha\over g^* m}.
$$
The equality arises  for $T_{\rm d}=T_{\rm eq} $. 

The main difference between the vorton and particle mechanisms is the presence of the $T_d/m_{\rm Pl}$ suppression factor in the
vorton case. The corresponding factor in the particle case is $\alpha/m$ which we can only bound to be less than unity unless we
specify the properties of the particle more exactly. In general decaying particles generate baryon asymmetry most efficiently if they decay at any time before they dominate the energy density of the Universe,
in contrast to decaying vortons which produce baryon asymmetry most efficiently if they  decay just as they come
to dominate the energy density. However, this comparison is slightly misleading; we have assumed a fixed
decay temperature and have varied the vorton formation temperature to maximise the baryon asymmetry, but the
 corresponding feature of the particle model, the dependence of $\alpha$ on $m$, is not specified in the general model. 
The feature that appears both in the vorton model and the general particle model is that entropy generated by the decay 
of objects that dominate the energy density reduces the net baryon asymmetry produced. In all cases baryon asymmetry is 
produced most efficiently if the object do not   dominate the energy density of the Universe before they decay. 
\vskip 15pt
\leftline {\bf 4. Discussion}

We have shown that remnants of superconducting strings, vortons, can decay
after a subsequent phase transition and these dissipating vortons can create
 baryon asymmetry. In the case of a GUT scale strings decaying at the electroweak scale, the resulting asymmetry
is not enough to explain observations. This is due to the fact that vortons
dominate the energy density of the Universe long before they decay.
Their decay results in a reheating of the Universe and an
 increase in the entropy density. This reheating is unlikely to have any effect on the
standard cosmology following the electroweak phase transition. If however there
was an intermediate transition, and the vortons never dominated the energy
density of the Universe, then their decay could explain the observed baryon
asymmetry of the Universe.

The question should be addressed as to how general the mechanism in this
letter is. In many GUT models there is a $\nu_R$ which acquires a mass at
the grand unified scale. If strings form when the GUT breaks, they
may acquire $\nu_R$ zero modes. In order to implement the 
`see-saw' mechanism the $\nu_R$ mixes with the standard neutrinos at the 
electroweak scale, resulting in a mass matrix with off-diagonal terms. 
Unless the electroweak Higgs field winds in the region of restored 
electroweak symmetry [9,11], then the $\nu_R$ will cease to be a zero mode 
after the electroweak phase transition, becoming  instead  a low-lying 
bound state. On the other hand, if the electroweak Higgs field winds
around the GUT string then the zero mode will remain. In addition, the
ordinary quarks and leptons will also become zero modes in the effective
electroweak string [12]. 

Neutrino zero modes appear in many GUT theories which produce 
strings, they  can be excited, leading to  neutral
currents flowing along the strings and consequently to the formation of vortons. 
Thus vortons will form in a wide class of GUT
models and the mechanism we have analysed is not restricted to the
specific case of SO(10) strings. 

This work shows that vorton remnants of superconducting strings are not
necessarily disastrous because they may not survive 
from the GUT scale down to nucleosynthesis, even if zero mode bearing vortons turn out to be
quantum mechanically stable. The model we have considered still has 
the Universe radiation dominated at nucleosynthesis and  is cosmologically
acceptable. Hence the existence or vortons is not enough to
rule out a theory,  one needs to check that the currents, and hence the vortons,
survive subsequent symmetry breakings. 

This mechanism could 
explain the observed baryon asymmetry depending on the scales of the 
phase transitions
leading to  vorton formation and vorton dissipation. There is one caveat to 
this in that any baryon asymmetry created by the dissipating vortons
could be erased by sphaleron processes at the electroweak scale. In the case
of SO(10), the asymmetry created is B+L preserving, and so is not erased
by electroweak processes, which only eliminate a B-L asymmetry. More generally,
many grand unified models create a B+L asymmetry, and so would evade 
sphaleron processes. The minimal SU(5) theory creates a B-L preserving 
asymmetry. 
However, it doesnot have cosmic strings, nor a $\nu_R$, so our mechanism is
not applicable in this case. Flipped SU(5) does have embedded defects [13]
and a $\nu_R$, so our mechanism may apply here if the defects are meta-stable.
However, even in the case of flipped SU(5) the baryon asymmetry produced 
is not necessarily destroyed by electroweak processes [14]. Some theories
with intermediate scale transitions have a $\nu_R$, and create a lepton
asymmetry. This is then converted to a by sphaleron processes at the 
electroweak scale [15]. Our analysis would be applicable to such theories.

We have also considered the late decay of any baryon number violating relic and
bounded the resulting asymmetry.
In our general analysis we have seen that the maximum 
asymmetry is produced if the relic decays before it dominates the
energy density of the Universe. If the decay occurs later than this, reheating and
entropy generation occur  and  the baryon asymmetry is diluted.

This work is supported in part by PPARC and the EU under HCM programme
(CHRX-CT94-0423). We wish to thank Stephen Davis for discussions.

\vskip 15pt

\leftline {\bf 5. References}
\noindent [1] W.B Perkins and A.C. Davis, Nucl Phys {\bf B406} (1993) 377
\vskip 9pt
\noindent [2] R. Brandenberger, B. Carter, A.C. Davis, and M. Trodden, hep-ph/9605382
\vskip 9pt
\noindent [3] R. Davis and E.P.S. Shellard, Nucl Phys {\bf B323} (1989) 209
\vskip 9pt
\noindent [4] B. Carter and X. Martin, Ann Phys {\bf 227} (1993) 151;
X. Martin and P. Peter, Phys Rev {\bf D51} (1995) 4092
\vskip 9pt
\noindent [5] E. Witten, Nucl Phys {\bf B249} (1985) 557
\vskip 9pt
\noindent [6] R. Brandenberger, A.C. Davis and M. Hindmarsh, Phys Lett {\bf B263} (1991)
239
\vskip 9pt
\noindent [7] A.C. Davis, M.A. Earnshaw and U.A. Wiedemann, Phys Lett {\bf B293} (1992)
123
\vskip 9pt
\noindent [8] R.L. Davis, Phys Rev {\bf D38} (1988) 3722
\vskip 9pt
\noindent [9] A.C. Davis and S.C. Davis, DAMTP/96-72, hep-ph/9608206 
\vskip 9pt
\noindent [10] C.T. Hill and L.M. Widrow, Phys Lett {\bf B189} (1987) 17;
M. Hindmarsh, Phys Lett {\bf B200} (1988) 429
\vskip 9pt
\noindent [11] A.C. Davis, S.C. Davis and W.B. Perkins (in preparation)
\vskip 9pt
\noindent [12] A.C. Davis and W.B. Perkins, Phys Lett (in press), DAMTP/96-71,
hep-ph/9610292
\vskip 9pt
\noindent [13] A.C. Davis and N.F. Lepora, Phys Rev {\bf D52} (1995) 7265
\vskip 9pt
\noindent [14] J. Ellis, D.V. Nanopoulos, K.A. Olive, Phys Lett {\bf B300}
(1993) 121
\vskip 9pt
\noindent [15] S.A. Abel and K.E.C. Benson, Phys Lett {\bf B335} (1994) 179
\end